\documentstyle[12pt,epsf]{article}
\newcommand{\newc}{\newcommand}
\newc{\gsim}{\lower.7ex\hbox{$\;\stackrel{\textstyle>}{\sim}\;$}}
\newc{\lsim}{\lower.7ex\hbox{$\;\stackrel{\textstyle<}{\sim}\;$}}
\newc{\gev}{\,{\rm GeV}}
\newc{\tev}{\,{\rm TeV}}
\newc{\ie}{{\it i.e.}}
\catcode`@=11
\long\def\@caption#1[#2]#3{\par\addcontentsline{\csname
  ext@#1\endcsname}{#1}{\protect\numberline{\csname
  the#1\endcsname}{\ignorespaces #2}}\begingroup
    \small
    \@parboxrestore
    \@makecaption{\csname fnum@#1\endcsname}{\ignorespaces #3}\par
  \endgroup}
\catcode`@=12

\begin{document}

\title{\small \rm \begin{flushright} \small{IASSNS-HEP 96/79}\\
\small{August 1996}\\
\end{flushright} \vspace{2cm}
\LARGE \bf Particle Physics for Cosmology\thanks{Invited
talk at `Critical Dialogues in Cosmology' conference celebrating the 250th
anniversary of Princeton University, June 1996.}\vspace{0.8cm}}
\author{Frank Wilczek\thanks{Research supported in part by DOE
grant DE-FG02-90ER40542.  E-mail:  wilczek@sns.ias.edu } \\
{\small \em School of Natural Sciences}\\
{\small \em Institute for Advanced Study}\\
{\small \em Princeton, NJ 08540}\\[.15in]} 

\begin{center}
\maketitle

\begin{abstract}

I discuss some central issues in particle physics which are
potentially relevant to cosmology.  I first briefly review the present
(glorious) experimental status of the Standard Model, emphasizing that
it provides a firm foundation both for early Universe cosmology and
for further exploration toward the basic laws of Nature.  I then
provide a critique, arguing that while there are no clear
discrepancies, there are several major, specific deficiencies of the
Standard Model which clearly point up its provisional character.  I
elaborate on the story theorists have made up to address one of these
problems, the problem of scattered multiplets, and show how upon
following it out one finds, within existing experiments, encouragement --
bordering on evidence -- for certain ambitious ideas regarding
unification and supersymmetry.  I briefly describe and contrast two
paradigms of supersymmetry breaking, which have markedly different
experimental and cosmological consequences.  I call attention to
specific connections with cosmology where appropriate throughout; and
near the end I make some more global
remarks.   Finally I venture a speculation suggesting, in a fairly
concrete way,
the possibility that
the laws of physics cannot,
in principle, be disentangled from cosmology.

\end{abstract}
\end{center}


\section{Triumph of the Standard Model}

The core of the Standard Model \cite{sm,weinsalam,sutheory} of
particle physics is easily displayed
in a single Figure, here Figure 1.  There are gauge groups
$SU(3)\times SU(2)\times U(1)$ for the strong, weak, and
electromagnetic interactions.  The gauge bosons associated with these
groups are minimally coupled to quarks and leptons according to the
scheme depicted in the Figure.  The non-abelian gauge bosons within
each of the
$SU(3)$ and $SU(2)$ factors also couple, in a canonical minimal form,
to one another.  The $SU(2)\times U(1)$ group is spontaneously broken
to the $U(1)$ of electromagnetism.  This breaking is parameterized in
a simple and (so far) phenomenologically adequate way by including an
$SU(3)\times SU(2)\times U(1)$ $(1, 2, -{1\over 2})$ scalar `Higgs'
field which condenses, that is has a non-vanishing expectation value
in the ground state.  Condensation occurs at weak coupling if the bare
(mass)$^2$ associated with the Higgs doublet is negative.

The fermions fall into five separate multiplets under
$SU(3)\times SU(2) \times U(1)$, as depicted in the Figure.
The color $SU(3)$ group acts horizontally; the weak $SU(2)$
vertically, and the hypercharges (equal to the average electric
charge) are as indicated.  Note that
left- and right-handed fermions of a single type generally transform
differently.  This reflects parity violation.  It also implies that
fermion masses, which of course connect the left- and right-handed
components, only arise upon spontaneous $SU(2)\times U(1)$ breaking.

Only one fermion family has been depicted in Figure 1; of course
in reality there are three repetitions of this scheme.
Also not represented are all the complications associated with the
masses and Cabibbo-like mixing angles among the fermions.
These masses and mixing angles are naturally accommodated
as parameters within the Standard Model, but I think it is fair to say
that they are not much related to its core ideas -- more on this below.

With all these implicit understandings and discrete choices, the core
of the Standard Model is specified by three numbers -- the universal
strengths of the strong, weak, and electromagnetic interactions.  The
electromagnetic sector, QED, has been established as an extraordinarily
accurate and fruitful theory for several decades now.  Let me now
briefly
describe the current status of the remainder of the Standard Model.

\begin{figure}
\centering
\epsfysize=3in
\vglue-.50in
\hglue0.75in\epsffile{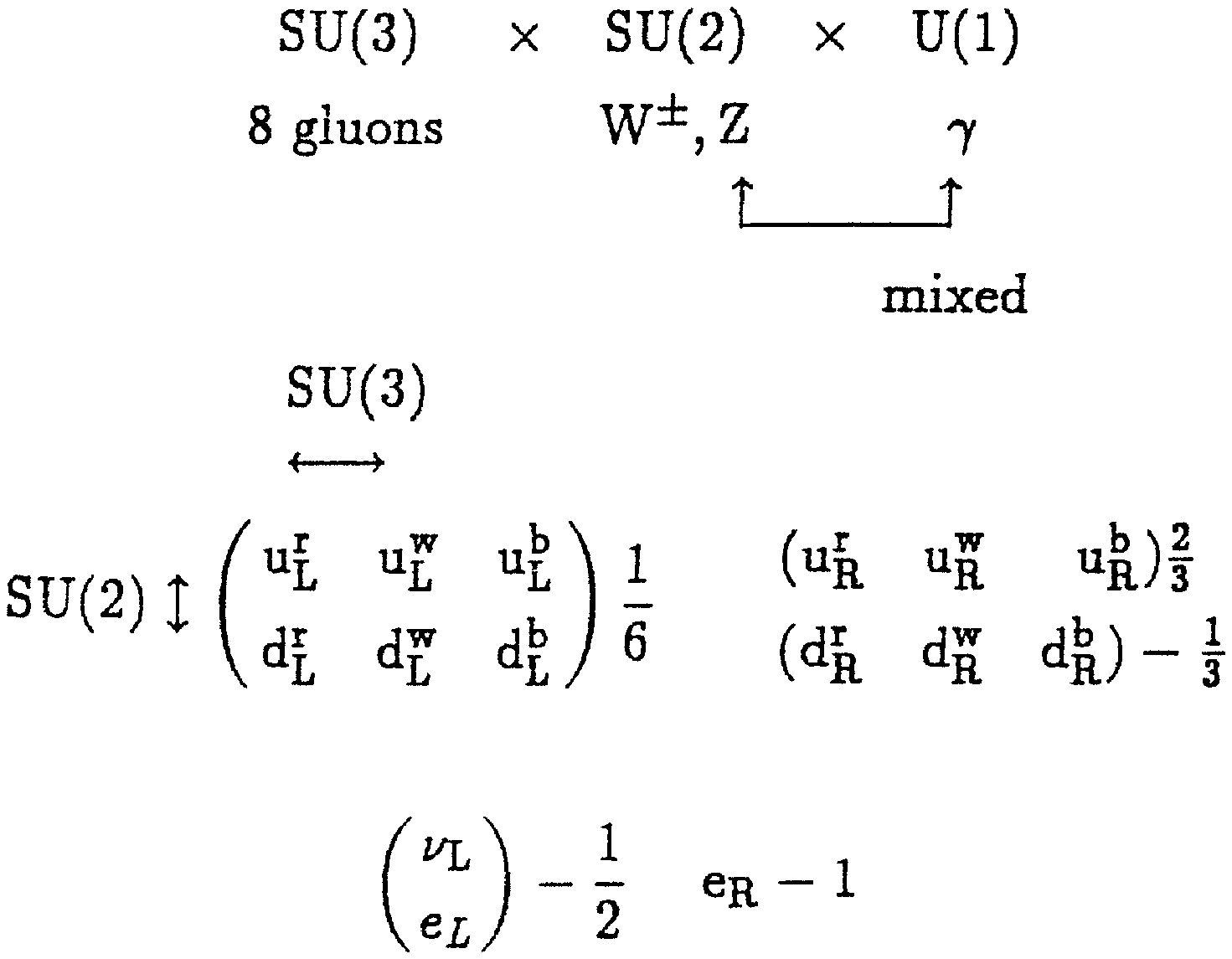}
\caption[]{The core of the Standard Model: the gauge groups and the
quantum numbers of quarks and leptons.  There are three gauge groups,
and five separate fermion multiplets (one of which, $e_R$, is a singlet).
Implicit in this Figure are the universal gauge couplings -- exchanges
of vector bosons -- responsible for the classic phenomenology of the
strong, weak, and electromagnetic interactions.
The triadic replication of quark and leptons, and the Higgs field whose
couplings and condensation
are responsible for $SU(2)\times U(1)$ breaking and for fermion masses and
mixings, are not indicated.}
\label{fig1}
\end{figure}

Some recent stringent tests of the electroweak sector of the
Standard Model are summarized in Figure 2.  In general each entry represents
a very different experimental arrangement,
and is meant to test a different fundamental
aspect of the theory, as described in the caption.  There is
precisely one parameter
(the mixing angle) available within the theory, to describe all these
measurements.
As you can see, the comparisons
are generally at the level of a per cent accuracy or so.
Over all, the agreement
appears remarkably good, especially to anyone familiar with the history
of weak interactions.  It is so good, that if the
small but significant residual discrepancies
in the branching ratios
$R_b$ and $R_c$ survive further scrutiny, one will certainly prefer
to interpret them as an indication of some peculiar
additional physics, rather
than as a breakdown of the fundamental principles behind the
$SU(2)\times U(1)$ theory.  Indeed, there are
some reasonably attractive concrete possibilities along
these lines \cite{bkm}.


\begin{figure}
\centering
\epsfysize=4in
\hspace*{0in}
\vglue-.7in
\hglue0.75in\epsffile{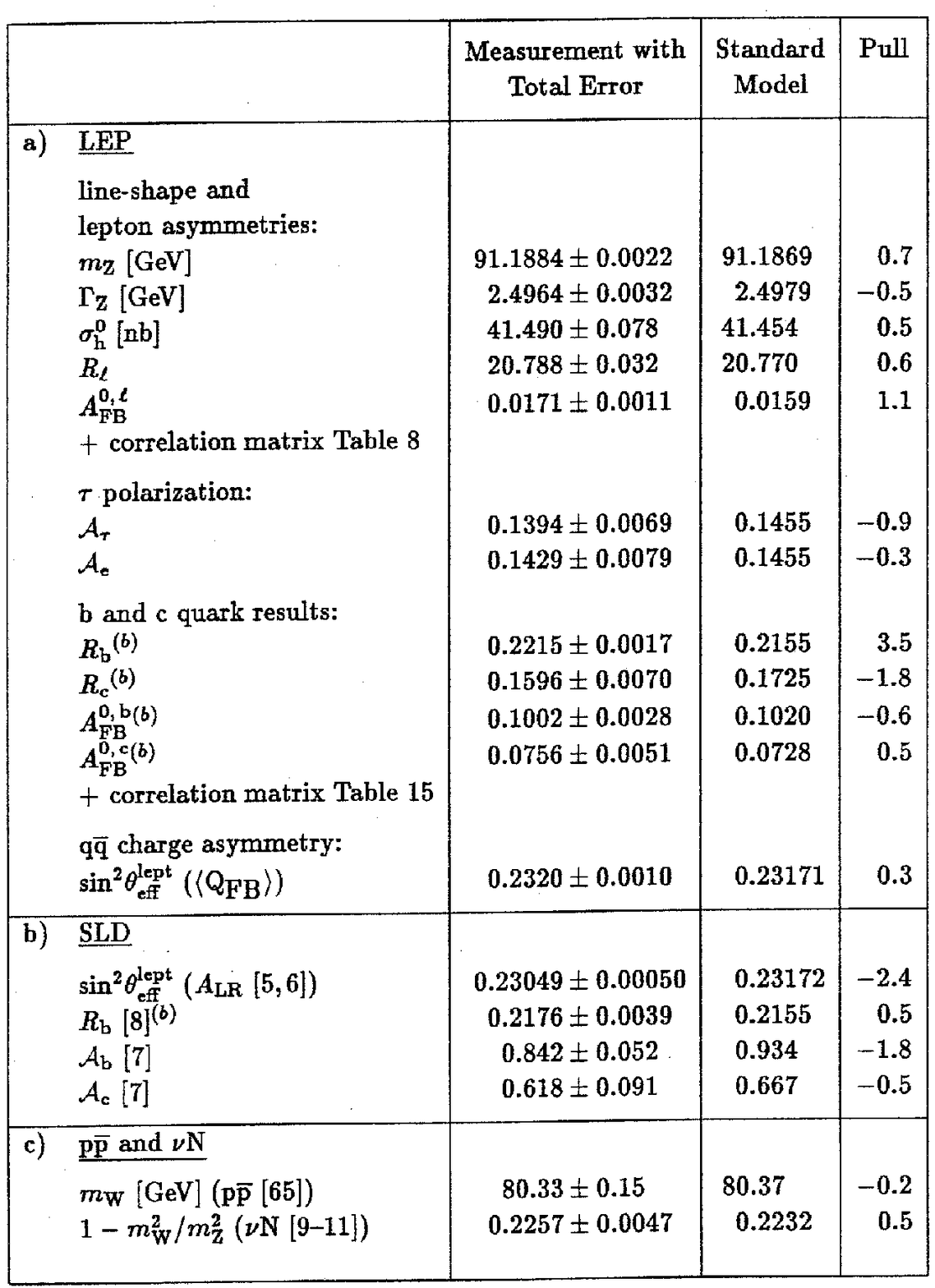}

\caption{A recent compilation of precision tests of
electroweak theory, from \protect\cite{ewfig} ,
to which you are referred for details.
Despite some `interesting' details, clearly the evidence for electroweak
$SU(2)\times U(1)$, with the simplest doublet-mediated symmetry breaking
pattern, is overwhelming.}
\label{fig2}
\end{figure}


Some recent stringent tests of the strong sector of the Standard Model are
summarized in Figure 3 \cite{qcdfig}.  Again a wide variety of very
different measurements 
are represented, as indicated in the caption.  A central feature of the
theory (QCD) is that the value of the coupling, as measured in different
physical processes, will depend in a calculable way upon the characteristic
energy scale of the process.  The coupling was predicted --
and evidently is now convincingly
measured -- to decrease as the inverse logarithm
of the energy scale: asymptotic
freedom.  Again, all the experimental results must be fit with just one
parameter -- the coupling at any single scale, usually chosen as
$M_Z$.  As you can see, the agreement between theory and
experiment is remarkably good.  The accuracy of
the comparisons is at the few per cent level.

\begin{figure}
\centering
\vglue-.65in
\epsfysize=3.5in
\hspace*{0in}
\epsffile{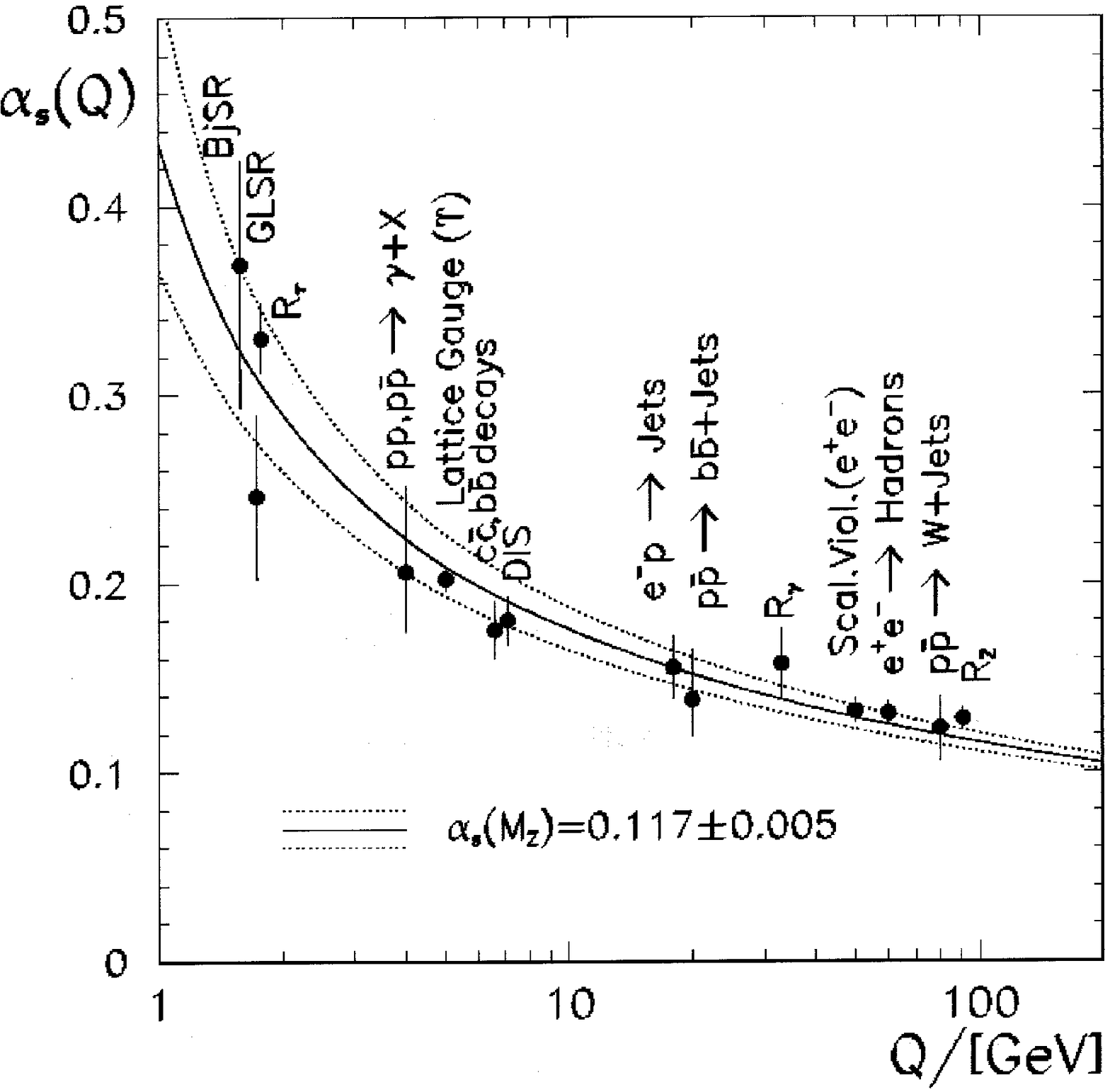}
\caption{A recent compilation of tests of QCD and asymptotic freedom, 
from \protect\cite{qcdfig},
to which you are referred for details.  Results are presented in the
form of determinations of the effective coupling $\alpha_s (Q)$ as a
function of the characteristic typical energy-momentum scale involved
in the process being measured.  Clearly the evidence for QCD in
general, and for the decrease of effective coupling with increasing
energy-momentum scale (asymptotic freedom) in particular, is
overwhelming.}
\label{fig3}
\end{figure}

Let me emphasize that these Figures barely begin to do justice to the
evidence for the Standard Model.  Several of the
results in them summarize quite
a large number of independent measurements, any
one of which might have falsified the theory.  For example, the single
point labeled `DIS' in Figure 3 describes literally hundreds of measurements
in deep inelastic scattering with different projectiles and
targets and at various energies and angles, which must
-- if the theory is correct --
all fit into a tightly constrained pattern.

The central theoretical principles of the Standard Model have
been in place
for nearly twenty-five years.
Over this interval the quality of
the relevant experimental data has become
incomparably better, yet no
essential modifications of these venerable principles has been required.
Let us now praise the Standard Model:
\bigskip

\noindent$\bullet$ The Standard Model is here to stay, and describes
{\it a lot}.

Since there is quite direct evidence for each of its fundamental
ingredients ({\it i.e}. its interaction vertices), and
since the Standard Model
provides an extremely economical packaging of these ingredients, I think
it is a safe conjecture that it will be used, for the foreseeable future,
as the working description of the phenomena within its domain.  And this
domain
includes a very wide range of phenomena --
indeed not only what Dirac called
``all of chemistry and most of physics''\footnote{Dirac was referring,
here, to quantum electrodynamics.}, but also the original problems of
radioactivity and nuclear interactions which inspired the birth of particle
physics in the 1930s, and much that was unanticipated.

\bigskip

\noindent$\bullet$ The Standard Model is a {\it principled\/} theory.

Indeed, its structure embodies a few basic principles:
special relativity, locality, and quantum mechanics,
which lead one to quantum
field theories, local symmetry (and, for the electro weaksector, its
spontaneous breakdown), and renormalizability (minimal coupling).
The last of these principles, renormalizability,
may appear rather technical and perhaps less compelling
than the others; we shall shortly have occasion to re-examine it
in a larger perspective.
In any case, the fact that the Standard Model is principled in this
sense is profoundly significant: it means that its predictions are
precise and unambiguous, and generally cannot be modified `a little bit'
except in very limited, specific ways.  This feature makes the
experimental success especially meaningful, since it becomes hard to
imagine that the theory could be approximately right without in some sense
being exactly right.

\bigskip

\noindent$\bullet$ The Standard Model {\it can be extrapolated}.

Specifically because of
the asymptotic freedom property,
one can extrapolate using the Standard Model from the observed domain
of experience to much larger energies and shorter distances.  Indeed, the
theory becomes simpler -- the fundamental interactions are all
effectively
weak -- in these limits.  The whole field of very early Universe cosmology
depends on this fact, as do the impressive
semi-quantitative indications for unification and supersymmetry I shall be
emphasizing momentarily.

\section{Deficiencies of the Standard Model}

Just because it is so comprehensive and successful, we should judge
the Standard Model by demanding criteria.
It is clearly an important part of the
Truth; the question becomes: How big a part?  Critical scrutiny reveals
several important shortcomings of the Standard Model:

\bigskip

\noindent$\bullet$ The Standard Model contains scattered multiplets with 
peculiar hypercharge assignments.

While little doubt can remain that the Standard Model is essentially
correct, a glance at Figure 1 is enough to reveal that it is not a
complete or final theory.  The
fermions fall apart into five lopsided pieces with peculiar hypercharge
assignments; this pattern needs to be explained.  Also the separate gauge
theories, which are quite mathematically and conceptually similar, are
fairly begging to be unified.

\bigskip

\noindent$\bullet$ The Standard Model supports the possibility of strong 
P and T violation \cite{hooft}.

There is a near-perfect match between the necessary `accidental' symmetries
of the Standard Model, dictated by its basic principles as enumerated above,
and the observed symmetries of the world.  The glaring exception is that
there is an allowed -- gauge invariant, renormalizable -- interaction which,
if present, would induce significant violation of the discrete symmetries
$P$ and $T$ in the strong interaction.  This is the notorious $\theta$ term.
$\theta$ is an angle which {\it a priori\/} one might expect to be of order
unity, but in fact is constrained by experimental limits on the neutron
electric dipole moment to be $\theta \lsim 10^{-8}$.  This problem can be
addressed by postulating a sort of quasi-symmetry 
(Peccei-Quinn \cite{pecci77} symmetry),
which roughly speaking corresponds to promoting $\theta$
to a dynamical variable -- a quantum field.  The quanta of this 
field \cite{weinberg78}, 
{\it axions}, provide an interesting dark matter candidate \cite{preskill83}.
Other possibilities for explaining the absence of strong $P$ and $T$ violation
have been proposed, but they require towers of hypotheses which
seem (to me) quite fragile.

I would like to emphasize, since it came up at the conference, that
in no way should the absence of strong $P$ and $T$ violation be taken
as evidence against QCD itself.  For practical purposes, one can take
the $\theta$ parameter to be fixed phenomenologically.  One finds it to
be very small -- and is done with it!

\bigskip

\noindent$\bullet$ The Standard Model does not address family problems.

There
are several distinct `family problems', ranging from the extremely
qualitative (digital) to the semi-qualitative
(some very small numbers and distinctive patterns) to the straightforward
but most challenging goal of doing full justice to
experience by computing (analog) experimental numbers with controlled,
small fractional errors:

Why are there three repeat families?  Rabi's famous question regarding
the muon -- ``Who ordered {\it that}?'' -- still has no convincing
answer.

How does one explain the very small electron mass?  The dimensionless
coupling associated with the electron mass, that is its strength of
Yukawa coupling to the Higgs field, is about $g_e \sim 2\times
10^{-6}$.  It is almost as small as the limits on the $\theta$ angle
(suggesting, perhaps, that strong P and T violation is just around the
corner?).  This question can of course be generalized -- all the
fermion masses, with one exception, are sufficiently small to beg
qualitative explanation.  The exception, of course, is the $t$ quark.
It is a fascinating possibility
that roughly the observed value of $t$ quark mass at low energies
might result after running from a wide range of fundamental couplings
at a high scale \cite{topquark}.
If so, one would have a satisfactory qualitative
explanation of the value of this parameter.

Why do the weak currents couple approximately in the order of masses?
That is, light with light, heavy with heavy, and intermediate with
intermediate.  Why are the mixings what they are -- small, but not
miniscule?  The same, for the CP violating phase in the weak currents
(parameterized invariantly by Jarlskog's $J$) \cite{jarlskog85}?
-- and, by the way, are we sure that $\theta \ll J$?
And so on ...

\bigskip

\noindent$\bullet$ The Standard Model does not allow non-vanishing
neutrino masses.

This is the only entry on my list that has a primarily {\it
experimental\/} motivation.  At present there are three quite
different experimental hints for non-vanishing neutrino masses: the
solar neutrino problem \cite{bahcall95}, the atmospheric neutrino 
problem \cite{lisi94},
and the Los Alamos oscillation experiment \cite{athan95}.
The Standard Model in its conventional
form does not allow non-zero neutrino masses.   However
I would like to
mention that only a very minimal extension of the theory is necessary
to accommodate such masses.  One can add a complete $SU(3)\times
SU(2)\times U(1)$ singlet fermion $N_R$ to the model.  $N_R$ can be
given, consistent with all symmetries and with the requirement of
renormalizability, a Majorana mass $M$.  Note especially that such a
mass does not violate $SU(2)\times U(1)$.  Likewise, $N_R$ can have a
Yukawa coupling to the to the ordinary lepton doublets through the
Higgs field.  Then condensation of the Higgs field activates the
``see-saw'' mechanism \cite{gell-mann79} to give a small mass for the 
observed neutrinos;
with $M \lsim 10^{15} {\rm Gev}$ a range of experimentally and perhaps
cosmologically interesting neutrino masses can be accommodated.

\bigskip

\noindent$\bullet$ Gravity is not included in the Standard Model.

This really represents (at least) two logically separate problems.

First there is the ultraviolet problem, the notorious
non-renormalizability of quantum gravity.  This provides an
appropriate context, in which to introduce the modern perspective
toward the whole concept of renormalizability.

Suppose that one were naive, and simply added the Einstein Lagrangian
for general relativity to the Standard Model, of course coupling the
matter fields appropriately (minimally).
Following Feynman and many others one
could derive, formally, the perturbation series for any physical
process.  One would find, however, that, the integrals over closed
loops generally diverges at the high-energy (ultraviolet) end.  Indeed
the graviton coupling has, in units where $\hbar = c =1$, dimensions
of $M_{\rm Planck}^{-1}$\footnote{This occurs because the kinetic
energy for the graviton arises from the Einstein action $\propto
M_{\rm Planck}^2\int \sqrt g R$ so that in expanding about flat space
one must take $g_{\mu \nu} = \eta_{\mu \nu} + {1\over M_{\rm Planck}}
h_{\mu \nu}$ in order to obtain a properly normalized
quadratic kinetic term for 
$h$. }.  Here $M_{\rm Planck} \approx 10^{19} ~{\rm Gev}~$ is a
measure of the stiffness of space-time.  Thus higher and higher order
terms will, on dimensional grounds, introduce higher and higher
factors of the ultraviolet cutoff to compensate.  However if we fix
(by notional experiments) the couplings at a given scale $\Lambda <<
M_{\rm Planck}$ and calculate corrections by only including
energy-momenta between the scale $P< \Lambda$ of interest and
$\Lambda$, the successive terms in perturbation theory will be
accompanied by positive powers of ${\Lambda \over M_{\rm Planck}}$ and
will be small.  Thus we can, for example, consistently set all
non-minimal couplings to zero at any chosen
energy-momentum scale well below
the Planck scale.  They will then be negligible for all practically
accessible scales.  For different choices
of the scale they will be different, but since
they are negligible this hardly matters.
This procedure is, in practice, the one we always adopt -- and the
Standard Model peacefully coexists with gravity, so long as we refuse
to consider $P \gsim M_{\rm Planck}$.

However from this perspective a second problem looms larger than ever.
The energy (and negative pressure) density of matter-free  space, the
notorious cosmological term, occurs as the coefficient $\lambda$
of the identity term in the action: $\delta {\cal L} =
\lambda \int \sqrt g$.  It has dimensions of (mass)$^4$, and on
phenomenological
grounds we must suppose $\lambda \lsim (10^{-12}~{\rm Gev}~)^4$.  The
question is: where does such a tiny scale come from?  What is so special
about the present state of the Universe, that the value of
the effective $\lambda$ for
it, which one might naively expect to reflect contributions from much
higher scales, is so effectively zeroed?

There may also be problems with forming a fully consistent quantum
theory even of low-energy processes involving black holes \cite{hawking76}.

\bigskip

Finally I will mention a question that I think has a rather different
status from the foregoing, although many of my colleagues would put it
on the same list, and maybe near the top:

The Standard Model begs the question ``{\it Why\/} does
$SU(2)\times U(1)$ symmetry break?''.

To me, this is an example of
the sort of metaphysical question that could easily fail
to have a meaningful answer.  There is absolutely nothing wrong, logically,
with the classic implementation of the Higgs sector as described earlier.
Nevertheless, one might well hunger for a wider context in which
to view the
existence of the Higgs doublet and its negative (mass)$^2$ -- or
a suitable alternative.

\section{Unification: Symmetry}

Each of the deficiencies of the Standard Model mentioned in the
previous section has provoked an enormous literature, literally hundreds
or thousands of papers.  Obviously I cannot begin to
do justice to all this work.
Here I shall concentrate on the first question, that of
deciphering the message of the scattered multiplets
and peculiar hypercharges.  Among our questions this one has
so far inspired the most concrete and compelling response --
a story whose
implications range far beyond the question which inspired it.


Given that the strong interactions are governed by transformations
among three color charges -- say RWB for red, white, and blue --
while the weak interactions are governed by transformations between
two others -- say GP for green and purple -- what
could be more natural than to embed both theories
into a larger theory of transformations among all five colors?
This idea has the additional attraction that an extra
U(1) symmetry commuting with the strong SU(3) and weak
SU(2) symmetries automatically appears,
which we can attempt to identify with the remaining gauge symmetry of
the Standard Model, that is
hypercharge.  For while in the separate SU(3) and SU(2) theories we
must throw out the two gauge bosons which couple respectively to
the color combinations R+W+B  and G+P, in the SU(5) theory we only
project out R+W+B+G+P, while the orthogonal
combination (R+W+B)-${3\over 2}$(G+P) remains.

Georgi and Glashow \cite{georgi74}
originated this line of thought, and showed how
it could be used to bring some order to the quark and lepton sector,
and in particular to
supply a satisfying explanation of the weird hypercharge assignments
in the Standard Model.  As shown in Figure 4, the five scattered
SU(3)$\times$SU(2)$\times$U(1) multiplets get organized into just two
representations of $SU(5)$.   It is an extremely non-trivial fact that
the known fermions fit so smoothly into $SU(5)$ multiplets.

\begin{figure}
\parskip=0pt
\underline{SU(5):  5 colors RWBGP}

$\underline{10}$: 2 different color labels (antisymmetric tensor)

$$\matrix{\rm u_L:&\rm RP,&\rm WP,&\rm BP\cr
\rm d_L:&\rm RG,&\rm WG,&\rm BG\cr
\rm u{^c_L}:&\rm RW,&\rm WB,&\rm BR\cr
&\rm (\bar B)&\rm (\bar R)&\rm (\bar W)\cr
\rm e{^c_L}:&\rm GP&&\cr
&(\ )&&\cr
}
\pmatrix{0&\rm u^c&\rm u^c&\rm u&\rm d\cr
&0&\rm u^c&\rm u&\rm d\cr
&&0&\rm u&\rm d\cr
&*&&0&\rm e\cr
&&&&0\cr}$$

$\underline{\bar 5}$: 1 anticolor label

$$\matrix{\rm d{^c_L}:&\rm \bar R,&\rm  \bar W,&\rm  \bar B\cr
\rm e_L:&\rm \bar P&&\cr
\nu_{\rm L}:&\rm \bar G&&\cr
}
\matrix{\rm \ \ \ \ \ \ \ \ \ (d^c&\rm d^c&\rm d^c&{\rm e}&\nu)\cr}$$
\def\boxtext#1{%
\vbox{%
\hrule
\hbox{\strut \vrule{} #1 \vrule}%
\hrule
}%
}
\centerline{
\vbox{\offinterlineskip
\hbox{\boxtext{\rm Y $= -{1\over 3}$ (R+W+B) $+{1\over 2}$ (G+P)}}
}}
\caption{Organization of the fermions in one family in $SU(5)$ multiplets.
Only two multiplets are required.  In passing from this form of
displaying the gauge quantum numbers to the form familiar in the
Standard Model, one uses the bleaching rules R+W+B = 0 and G+P = 0 for
$SU(3)$ and $SU(2)$ color charges (in antisymmetric combinations).
Hypercharge quantum numbers are identified using the formula in the
box, which reflects that within the larger structure $SU(5)$ one only
has the combined bleaching rule R+W+B+G+P = 0.  The economy of this
Figure, compared to Figure 1, is evident.}
\end{figure}

In all the most promising unification schemes,
what we ordinarily think of as matter and anti-matter appear on
a common footing.  Since the fundamental
gauge transformations do not alter the chirality of fermions,
in order to
represent the most general transformation possibilities
one should use fields of
one chirality, say left, to represent the fermion degrees of
freedom.   To do this, for a given fermion, may require a charge
conjugation operation.
Also, of course, once we contemplate changing strong
into weak colors it will be difficult to prevent quarks and leptons
from appearing together in the same multiplets.
Generically, then,
one expects that in unified theories it will not be possible
to make a global distinction between matter and anti-matter and that
both baryon number $B$ and lepton number $L$ will be violated, as
they definitely are in $SU(5)$ and its extensions.

As shown in Figure 4, there is one group of ten
left-handed fermions that have
all possible combinations of one unit of each of two different colors, and
another group of five left-handed fermions that each carry
just one negative unit of some
color.  (These are the ten-dimensional antisymmetric tensor and the
complex conjugate of the five-dimensional vector
representation, commonly  referred to as the ``five-bar''.)  What is
important for you
to take away from this discussion
is not so much the precise details of the scheme,
but the idea that {\it the structure of the Standard Model, with the
particle assignments gleaned from decades of experimental effort and
theoretical interpretation, is perfectly reproduced by a
simple abstract set
of rules for manipulating symmetrical symbols}.  Thus for example
the object
RB in this Figure has just the strong, electromagnetic, and
weak interactions we expect of the complex
conjugate of the right-handed up-quark, without our having to instruct
the theory further.  If you've never done it I heartily recommend
to you the
simple exercise of working out the hypercharges of the objects in
Figure 4 and checking against what you
need in the Standard Model
-- after doing it, you'll find it's impossible
ever to look at the standard model
in quite the same way again.

Although it would be inappropriate to elaborate the necessary group theory
here, I'll mention that there is a beautiful extension of $SU(5)$ to
the slightly larger group $SO(10)$, which permits one to unite
all the fermions
of a family into a single multiplet \cite{georgi75}.  In fact the relevant
representation for the fermions is a 16-dimensional spinor representation.
Some of its features are depicted in Figure 5.  The 16th member of a family
in $SO(10)$, beyond the 15 familiar degrees of freedom with a Standard Model
family, has the quantum numbers of the right-handed neutrino $N_R$ as
discussed above.   This emphasizes again
how easy and natural is the extension
of the Standard Model to include neutrino masses using the see-saw mechanism.

\begin{figure}
\parskip=0pt
\hglue0.75in\underline{SO(10): 5 bit register}
\vglue-.15in
$$(\pm \pm \pm \pm \pm)\ \ :\ \ \underline{\rm even}\ \  \# \  of\  -$$
$$10:\matrix{(++-|+-)&6&\rm (u_L,d_L)\cr
(+--|++)&3&\rm u{^c_L}\cr
(+++|--)&1&\rm e{^c_L}\cr}$$

$$\bar 5:\matrix{(+--|--)&\bar 3&\rm d{^c_L}\cr
(---|+-)&\bar 2&{\rm (e_L},\nu_L)\cr}$$
\nopagebreak
$$1:\matrix{(+++|++)&1&\rm N_R\cr}$$

\caption{Organization of the fermions in one family, together with a
right-handed neutrino degree of freedom, into a single multiplet under
$SO(10)$.  The components of the irreducible spinor representation,
which is used here, can be specified in a very attractive way by using
the charges under the $SO(2)\otimes SO(2)\otimes SO(2)\otimes
SO(2)\otimes SO(2)$ subgroup as labels.  They then appear as arrays of
$\pm$ signs, resembling binary registers.  There is the rule that one
must have an even number of - signs.  Strong $SU(3)$ acts on the first
three components, weak $SU(2)$ on the final two.  The $SU(5)$ quantum
numbers are displayed in the left-hand column, the number of entries
with each sign-pattern just to the right, and finally the usual
Standard Model designations on the far right.}
\end{figure}

\section{Unification: Dynamics, and a Big Hint of \break 
Supersymmetry\protect\cite{unification}}

\subsection{The Central Result}

We have seen that simple unification schemes are successful at the
level of
{\it classification}; but new questions arise when we consider the
dynamics which underlies them.

Part of the power of gauge symmetry is that it fully dictates the
interactions of the gauge bosons, once an overall coupling constant
is specified.  Thus if SU(5) or some higher symmetry were exact, then
the fundamental
strengths of the different color-changing interactions would have
to be equal, as would the
(properly normalized) hypercharge coupling strength.  In reality the
coupling strengths of the gauge bosons in SU(3)$\times$SU(2)$\times$U(1)
are observed not to be equal, but rather to follow the pattern
$g_3 \gg g_2 > g_1$.

Fortunately, experience with QCD emphasizes that couplings ``run''.
The physical mechanism of this effect is that in quantum field theory
the vacuum must be regarded as a polarizable medium, since virtual
particle-anti-particle pairs can screen charge.  Thus one might expect
that effective charges measured at shorter distances, or equivalently
at larger energy-momentum or mass scales, could be different from what
they appear at longer distances.  If one had only screening then the
effective couplings would grow at shorter distances, as one penetrated
deeper insider the screening cloud.  However it is a famous fact 
\cite{sutheory} that due to paramagnetic
spin-spin attraction of like charge vector gluons \cite{nielsen81},
these particles tend
to {\it antiscreen\/} color charge, thus giving rise to
the opposite effect -- asymptotic freedom --
that the effective coupling tends to shrink at short distances.  This
effect is the basis of all perturbative QCD phenomenology, which is a vast
and vastly successful enterprise, as we saw
in Figure 3.

For our present purpose of understanding
the disparity of the observed couplings, it is just what the doctor
ordered.
As was first pointed out by Georgi, Quinn, and Weinberg \cite{quinn74},
if a
gauge symmetry such as SU(5) is spontaneously broken at some very short
distance then we should not expect that the effective couplings probed at
much larger distances, such as are actually measured at practical
accelerators, will be equal.  Rather they will all have have been affected
to a greater or lesser extent by vacuum screening and anti-screening,
starting from a common value at the unification scale but then diverging
from one another at accessible accelerator scales.
The pattern $g_3 \gg g_2 > g_1$ is just what one should
expect, since the antiscreening or asymptotic freedom effect is more
pronounced for larger gauge groups, which have more types of virtual
gluons.

The marvelous thing is that the running of the couplings
gives us a truly quantitative
handle on the ideas of unification, for the following reason.  To fix
the relevant aspects of unification, one basically needs
only to fix two parameters: the scale at which the couplings unite, which
is essentially the scale at which the unified symmetry breaks; and
their value when then unite.  Given these, one calculates three outputs:
the three {\it a priori\/} independent couplings for the gauge groups
SU(3)$\times$SU(2)$\times$U(1) of the Standard Model.
Thus the framework is eminently
falsifiable.  The miraculous thing is, how close it comes to working
(Figure 6).

\begin{figure}
\centering
\epsfysize=3.5in
\hspace*{0in}
\vglue-0.75in
\hglue0.50in\epsffile{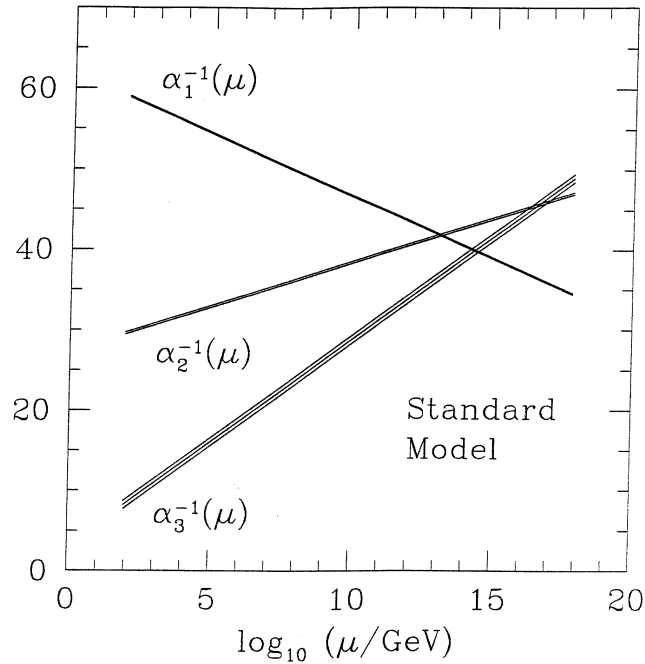}
\caption{Evolution of Standard Model effective (inverse) couplings toward small
space-time distances, or large energy-momentum scales.  Notice that
the physical behavior assumed for this Figure is the direct
continuation of Figure 3, and has the same conceptual basis.  The
error bars on the experimental values at low energies are reflected in
the thickness of the lines.  Note the logarithmic scale.  The
qualitative aspect of these results is extremely encouraging for
unification and for extrapolation of the principles of quantum field
theory, but there is a definite small discrepancy with recent
precision experiments.}
\label{fig6}
\end{figure}

The unification of couplings occurs at a very large mass scale,
$M_{\rm un.} \sim 10^{15}~{\rm Gev}$.  In the simplest version, this
is the magnitude of the scalar field vacuum expectation value that
spontaneously breaks SU(5) down to the
standard model symmetry SU(3)$\times$SU(2)$\times$U(1),
and is analogous to the scale $v \approx 250~ {\rm
Gev}$ for electroweak symmetry breaking.  The largeness of
this large scale mass scale
is important in several ways.

$\bullet~$ It
explains why the exchange of gauge bosons that are in SU(5) but not
in SU(3)$\times$SU(2)$\times$U(1), which re-shuffles
strong into weak colors
and generically violates the conservation of baryon number,
does not lead to a catastrophically quick decay of nucleons.  The rate
of decay goes as the inverse fourth power of the mass of the exchanged
gauge particle, so the baryon-number violating processes are predicted to
be far slower than ordinary weak processes, as they had better be.

$\bullet~$ $M_{\rm un.}$ is significantly smaller than the Planck scale
$M_{\rm Planck} \sim 10^{19}~{\rm Gev}$ at which exchange of gravitons
competes quantitatively with the other interactions, but not ridiculously
so.  This indicates that while the unification of couplings calculation
itself is probably safe from gravitational corrections, the unavoidable
logical next step in unification must be to bring gravity into the mix.

$\bullet~$ Finally one must ask how the tiny ratio of
symmetry-breaking mass scales $v/M_{\rm un.} \sim 10^{-13}$
required arises dynamically, and whether it is stable.  This is the
so-called gauge hierarchy problem, which we shall discuss in a
more concrete
form a little later.

The success of the GQW calculation in
explaining the observed hierarchy $g_3 \gg g_2 > g_1$ of
couplings and the approximate stability of the proton is quite
striking.
In performing it, we assumed that the known and
confidently expected particles of the Standard Model exhaust
the spectrum up to the unification scale, and that the
rules of quantum field
theory could be extrapolated without alteration
up to this mass scale -- thirteen orders
of magnitude beyond the domain they were designed to describe.
It is a triumph for minimalism, both existential and conceptual.


However, on further examination
it is not quite good enough.  Accurate modern measurements
of the couplings show a small but definite discrepancy between the
couplings, as appears in Figure 6.  And heroic dedicated experiments to
search for proton decay did not find it \cite{blewitt85}; they currently
exclude the minimal SU(5) prediction
$\tau_p \sim 10^{31}~{\rm yrs.}$ by about two orders of magnitude.

Given the scope of the extrapolation
involved, perhaps we should not have
hoped for more.
There are several perfectly plausible bits of physics
that could upset the calculation, such as the existence of particles
with masses much higher than the electroweak but much smaller than the
unification scale.  As virtual particles these would affect the running
of the couplings, and yet one
certainly cannot exclude their existence on direct experimental
grounds.  If we just add particles in some haphazard
way things will
only get
worse: minimal SU(5) nearly works, so the generic perturbation
from it will be deleterious.  This is a major difficulty for so-called
technicolor models, which postulate many new particles in
complicated patterns.
Even if some {\it ad hoc\/}
prescription could be made to work,
that would be a disappointing outcome from what
appeared to be one of our most precious, elegantly
straightforward clues regarding physics well
beyond the Standard Model.

Fortunately, there is a theoretical idea which is attractive in many
other ways, and seems to point a way out from this impasse.  That is
the idea of supersymmetry \cite{ferrara86}.  Supersymmetry is a symmetry that 
extends
the Poincare symmetry of special relativity
(there is also a general relativistic version).  In a supersymmetric
theory one has not only
transformations among particle states with different energy-momentum but
also between particle states of different {\it spin}.  Thus spin 0
particles can be put in multiplets together with spin ${1\over 2}$
particles, or spin ${1\over 2}$ with spin 1, and so forth.

Supersymmetry is certainly not a symmetry in nature: for example, there
is certainly no bosonic particle with the mass and charge of the electron.
More generally if one defines the $R$-parity quantum number
$$
R~\equiv~ (-)^{3B+L+2S}~,
$$
which should be accurate to the extent that baryon and lepton number are
conserved, then one finds that all currently known particles are
$R$ even whereas their supersymmetric partners would be $R$ odd.
Nevertheless there are
many reasons to be interested in supersymmetry, and especially in
the hypothesis that supersymmetry is effectively broken at a relatively
low scale, say $\approx$ 1 Tev.  Anticipating this for the moment, let
us consider the consequences for running of the couplings.

The effect of low-energy supersymmetry on the running of the couplings
was first considered long ago \cite{dimopoulos81},
well before the discrepancy described above
was evident experimentally.
One might
have feared that such a huge expansion of the theory, which essentially
doubles the spectrum, would utterly destroy the approximate success of
the minimal SU(5) calculation.  This is not true, however.  To a first
approximation since supersymmetry is a space-time rather than an internal
symmetry it does not affect the group-theoretic structure of the
calculation.  Thus to a first approximation the absolute
rate at which the couplings run
with momentum is affected, but not the relative rates.  The main effect
is that the supersymmetric partners of the color gluons, the gluinos,
weaken the asymptotic freedom of the strong interaction.  Thus they
tend to
make its effective
coupling decrease and approach the others more slowly.  Thus
their merger requires a longer lever arm,
and the scale at which the couplings meet increases by an order of
magnitude or so, to about 10$^{16}$ Gev.
Also the common value of the effective couplings at
unification is slightly larger than in conventional unification
(${g^2_{\rm un.} \over 4\pi } \approx {1\over 25}$ {\it versus\/}
${1\over 40}$).  This increase in unification scale
significantly reduces the predicted rate for proton decay
through exchange of the dangerous color-changing gauge bosons,
so that it no longer conflicts with existing
experimental limits.

Upon more careful examination there is another effect of low-energy
supersymmetry on the running of the couplings, which although quantitatively
small has become of prime interest.  There is an important exception to
the general rule that adding supersymmetric partners does not immediately
(at the one loop level)
affect the relative rates at which the couplings run.  This
rule works for particles that come in complete SU(5) multiplets, such as
the quarks and leptons (which, since they don't
upset the full SU(5) symmetry, have basically no effect)
or for the supersymmetric partners of the
gauge bosons, because they just renormalize the existing,
dominant effect of the
gauge bosons themselves.  However there is one peculiar additional
contribution, from the supersymmetric partner of the Higgs doublet.
It affects only the weak SU(2) and hypercharge U(1) couplings.
(On phenomenological grounds the
SU(5) color triplet partner of the Higgs doublet must be extremely massive,
so its virtual exchange is not important below the unification scale.
{\it Why\/}
that should be so, is another aspect of the hierarchy problem.)
Moreover, for slightly technical reasons even in the
minimal supersymmetric model it is necessary to have
two different Higgs doublets with opposite hypercharges\footnote{Perhaps
the simplest, though not the
most profound, way to appreciate the reason for this has to do with
anomaly cancelation.  The minimal
spin-1/2 supersymmetric partner of the
Higgs doublet is chiral and has non-vanishing hypercharge, introducing
an anomaly.  By including a partner for the anti-doublet, one cancels
this anomaly.}.
The net effect of
doubling the number of Higgs fields and including their supersymmetric
partners is a sixfold enhancement of the asymmetric
Higgs field contribution to
the running of weak and hypercharge couplings.  This causes a
small, accurately calculable change in the calculation.
{}From Figure 7 you see that it is a most welcome one.   Indeed,
in the minimal
implementation of supersymmetric unification, it puts the running of
couplings calculation right back on the money \cite{ellis91}.

\begin{figure}
\centering
\epsfysize=3.5in
\hspace*{0in}
\vglue-.75in
\hglue0.65in\epsffile{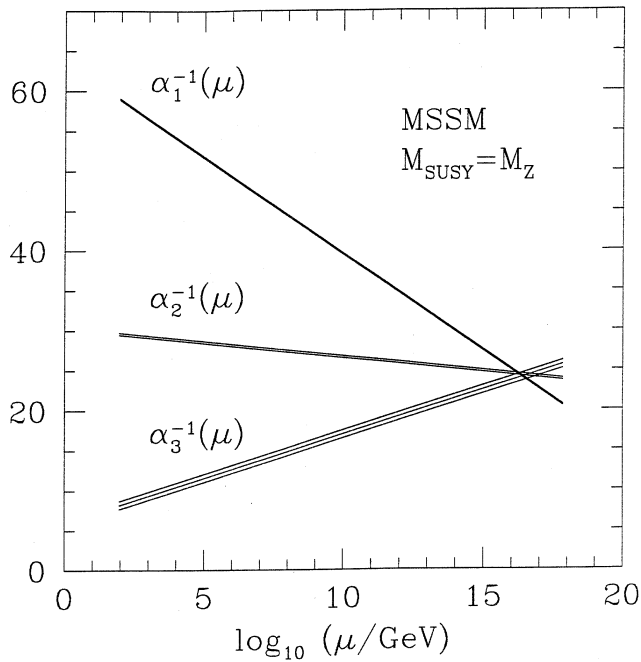}
\caption{Evolution of the effective (inverse) couplings in the minimal
extension of the Standard Model, to include supersymmetry.  The
concepts and qualitative behaviors are only slightly modified from
Figure 6 (a highly non-trivial fact!) but the quantitativive result is
changed, and comes into adequate agreement with experiment.  I would
like to emphasize that results along these lines were published well
before the difference between Figure 6 and Figure 7 could be resolved
experimentally, and in that sense one has already derived a successful
{\it prediction\/} from supersymmetry.}
\label{fig7}
\end{figure}

Since the running of the couplings with scale is logarithmic the
unification of couplings calculation is not
terribly
sensitive to the exact scale at which supersymmetry is broken,
say between 100 Gev and 10 Tev.  (To avoid confusion later let me
\break\hfil 
specify that at this point what I mean by
the scale at which
supersymmetry is broken is the characteristic mass splitting between
supersymmetric partners in a minimal extension of the Standard Model.
Later we shall use the term in a different sense, as referring to
the largest splitting between supersymmetric partners in the entire
world-spectrum; this could be much larger, and indeed in popular models
it almost invariably is.  This ambiguous terminology is endemic in the
literature; fortunately, the meaning is usually clear from the context.)
There have
been attempts to push the calculation further, in order
to address this question of
the supersymmetry breaking scale, but they are controversial.
For example, comparable uncertainties arise from the
splittings among the
very large number of particles with masses of order the unification scale,
whose theory is poorly developed and unreliable.  Superstring theory
suggests \cite{dienes96} many possible ways in which the simple calculation
described here might go wrong\footnote{Indeed,
in the simplest superstring-inspired models
it is not entirely easy to
accommodate the `low' value of the unification scale compared to the
Planck scale.};
if we take the favorable result of this calculation
at face value, we must conclude that
none of them happen.

In any case, if we are not too greedy
the main points still shine through:

$\bullet$ If supersymmetry is to fulfill its destiny
of elucidating the hierarchy problem in a straightforward way,
then the supersymmetric partners of
the known particles cannot be much heavier than the SU(2)$\times$U(1)
electroweak breaking scale, \ie\ they
should not be beyond the expected reach of LHC.

$\bullet$ If we
assume this to be the case
then the meeting of the couplings takes place in
the simplest minimal models of unification, to adequate accuracy,
without further assumption.  This is
a most remarkable and non-trivial fact.

\subsection{Implications}

The preceding result, taken at face value, has extremely profound
implications:

$\bullet$ Quantum field theory, and specifically its characteristic
vacuum polarization effects leading to asymptotic freedom and running
of the couplings, continue to
work quantitatively up to energy scales many orders
of magnitude beyond where they were discovered and established.

I would like to emphasize also
some negative implications of this: there are things
that might
have, but do {\it not}, happen.  It might have happened that the known
particles are some complicated composites of more elementary objects, or
that many additional
strong couplings appeared at higher energies (technicolor), or that
additional dimensions became dynamically active,
or that particle physics simply dissolved into some amorphous mess.
Unless Figure 7 is a cruel joke on the part of mother Nature, none of
this happens, or at least the complications are in a strong, precise sense
walled off from the Standard Model and the dynamical evolution
of its couplings.

$\bullet$ Supersymmetry, in its virtual form, has already been discovered.

This provides, in my opinion, a very good
specific reason to be hopeful about
the future of experimental
particle physics at the high-energy frontier.

\section{More About Supersymmetry\protect\cite{ferrara86}}

\subsection{Why Supersymmetry is a Good Thing}

\ \ \ \  $\bullet$ You will notice that we have made progress in uniting the
gauge bosons with each other, and the various quarks and leptons with each
other, but not the gauge bosons with the quarks and leptons.  It takes
supersymmetry -- perhaps spontaneously broken -- to make this feasible.

$\bullet$ Supersymmetry was invented in the context of string theory, and
seems to be necessary for constructing consistent
string theories containing gravity
(critical string theories) that are at all realistic.

$\bullet$ Most important for our purposes, supersymmetry can help
us to understand
the vast disparity between weak and unified symmetry breaking
scales mentioned above.  This disparity is known as the
gauge hierarchy problem.  It actually raises several distinct
problems,
including the following.
In calculating radiative corrections to the
(mass)$^2$ of the Higgs particle from diagrams of the type shown
in Figure 8
one finds an
infinite, and also large, contribution.  By this I mean that
the divergence is quadratic in the ultraviolet cutoff.  No ordinary
symmetry will make its coefficient vanish.
If we imagine that the unification scale provides the cutoff, we find
that the radiative correction to the (mass)$^2$ is much larger than the
final value we want.
(If the Higgs field were
composite, with a soft form factor, this problem might be ameliorated.
Following that road leads to technicolor,
which as mentioned before seems to lead
us far away from our best source of inspiration.)
As a formal matter one can simply cancel the radiative
correction against a large
bare contribution of the opposite sign, but in the absence of some
deeper motivating
principle this seems to be a horribly ugly procedure.
Now in a supersymmetric theory for any set of virtual particles circulating
in the loop there will also be another graph
with their supersymmetric partners circulating.  If the partners were
accurately degenerate, the contributions would cancel.  Otherwise, the
threatened quadratic divergence will be cut off only at virtual momenta
such
that the difference in (mass)$^2$ between the virtual
particle and its supersymmetric partner is relatively
negligible.  Thus we will
be assured adequate cancelation if and
only if supersymmetric partners are not
too far split in mass -- in the present context, if the splitting is not
much greater than the weak scale.  This is (a crude version of) the most
important {\it quantitative\/} argument which suggests the relevance
of ``low-energy'' supersymmetry.

\begin{figure}
\centering
\vglue-.75in
\epsfysize=3.5in
\hspace*{0in}
\epsffile{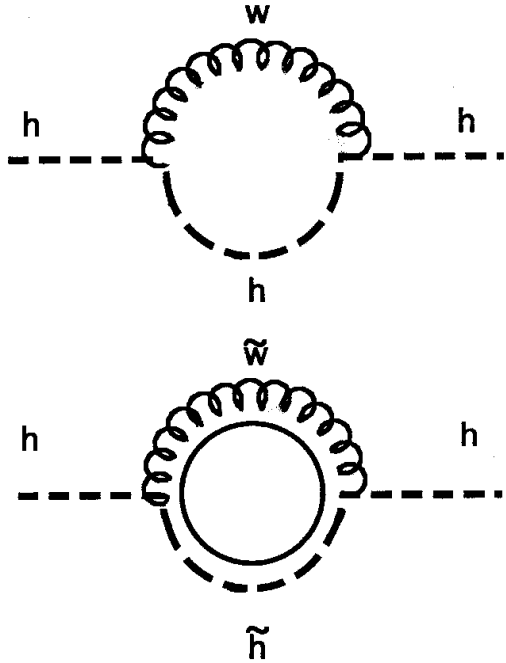}
\caption{Contributions to the Higgs field self-energy.  These graphs give
contributions to the Higgs field self-energy which separately are
formally quadratically divergent, but when both are included the
divergence is removed.  In models with broken supersymmetry a finite
residual piece remains.  If one is to obtain an adequately small
finite contribution to the self-energy, the mass difference between
Standard Model particles and their superpartners cannot be too great.
This -- and essentially only this -- motivates the inclusion of
virtual superpartner contributions in Figure 7 beginning at relatively
low scales.}
\label{fig8}
\end{figure}

$\bullet$  Supersymmetric field theories have
many special features, which make them especially interesting,
and perhaps promising, phenomenologically.

I cannot be very specific about this here, both because there are as yet
no canonical models and because the subject
is excessively technical, but let
me just mention some appropriate concepts: radiative $SU(2)\times U(1)$
breaking associated with the heavy top quark; doublet-triplet splitting
mechanisms; approximate
flat directions for generating large mass hierarchies.
Supersymmetric models also have additional mechanisms for neutral
flavor-changing processes and CP violation, which in generic models
are dangerously large, but in appropriate models can be suppressed down
to
a level which is interesting -- but not {\it too\/} interesting --
experimentally.

\subsection{Alternative Paradigms of Supersymmetry \break 
Breaking\protect\cite{kolda96}}

Theoretical control over the many
new parameters that arise even in minimal supersymmetric extensions of
the standard model is not adequate to guide one through a bewildering
number of choices that substantially affect the predictions for
observable phenomena.  

Recently a class of models that potentially enjoy much greater
predictive power has been identified \cite{models}.  
The leading idea of these
models is that the fundamental breaking of supersymmetry occurs in a
``hidden sector'' and is characterized by a scale 
$\Lambda \lsim 100\tev$.
This breaking is supposed to be conveyed to our observable
sector, including the minimal supersymmetric Standard Model, by a
specific, highly symmetric 
interaction term\footnote{
In the minimal messenger model (MMM) supersymmetry breaking is conveyed to
the standard model particles by the matter fields 
$(q,l)+(\overline{q},\overline{l}$) having the
quantum numbers ${\bf 5}+{\bf \overline{5}}$ under $SU(5)$.  These
fields are assumed to acquire supersymmetry breaking masses 
through  their couplings to a standard
model singlet field $X$ which develops vacuum expectation values along its
scalar as well as its $F$--components.  
Explicit models where this occurs dynamically have been constructed 
\cite{models}.  
Typically these models contain the superpotential couplings
\begin{equation}
W = \lambda_q \overline{q}qX+\lambda_l \overline{l}lX
\label{eq1}
\end{equation}
which generates a non--supersymmetric spectrum for the $(q,l)$ fields
through $F_X \neq 0$.  The $(q,l)$ fields, through their standard
model gauge interactions, induce SUSY breaking masses for the R-odd
scalars $(\widetilde{q}, \widetilde{\ell}$) and gauginos 
($\widetilde{B}, \widetilde{W},\widetilde{g})$.}.
This is to be contrasted with the situation in a
class of models, commonly referred to as  
supergravity models, that has been much more popular.  In supergravity
models  
the fundamental symmetry breaking scale is 
$\Lambda \sim 10^9\tev$, and the hidden sector communicates with
the observable sector by gravitational-strength interactions.  A
decisive phenomenological difference between these classes of models
arises because of the different role played by the 
gravitino $\widetilde G$, which is 
essentially the Nambu-Goldstone fermion of
spontaneous supersymmetry breaking.
Its
mass is generically given as 
$M_{\widetilde G} = {\Lambda^2\over M_{\rm Planck}}$, whereas its couplings
to ordinary matter scale as $1\over \Lambda^2$ \cite{fayet}.

In supergravity
models the gravitino is massive and very weakly coupled, and 
although it could be significant cosmologically, it plays no very
direct role in ordinary phenomenology.  In the
messenger models with a (relatively)
much lower supersymmetry breaking
scale the gravitino is exceedingly light, and
although its couplings are very feeble they can be such as to cause
the next lightest, photino-like particle $\widetilde {N_1} $ to decay via
$\widetilde {N_1} \rightarrow \gamma + \widetilde G$ within a distance that
might be difficult to resolve experimentally \cite{signals}.  
Generically, then, in
these models one will find that pair-production of supersymmetric
(R-odd) particles will lead eventually, perhaps after a cascade of
decays, to final states containing $\gamma \gamma$ and missing 
energy \cite{signals}.
The recent observation \cite{cdf} of a dramatic event with a final state
containing hard  $e^+e^-\gamma \gamma$ and missing transverse energy by
the CDF collaboration therefore 
lends special interest to these models.

One can work out the consequences
of a minimal implementation of a low-energy messenger
model for supersymmetry breaking \cite{kolda96}.  Outstanding features
of 
this model
include: generic production of final states including $\gamma \gamma$
and missing energy; 
clean separation of scales in the R-parity odd spectrum, with the
squarks and gluinos much heavier than the electroweak 
gauginos and sleptons; and
very small mixing of the $SU(2)\times U(1)$ neutralinos with Higgsinos
and with each other.  Many
specific features are predicted in terms of very few parameters, so
that the model could be rigorously tested, or eliminated, with quite a
small amount of relevant data.  If the CDF 
$e^+e^-\gamma \gamma$ event represents a first indication, 
such data is not far beyond current reach.


Of course in insisting upon an absolutely minimal model we may
be over reaching.  Some apparent difficulties with the
minimal model are removed upon careful treatment of coupling
renormalization and proper inclusion of D-terms;
nevertheless, one must remain open to such possibilities as different
overall mass scales for the sleptons as against the gauginos, or
additional contributions to the parameters of the Higgs sector.


Finally let me briefly mention another possibility for physics capable
of 
generating the CDF $e^+e^-\gamma \gamma$ event, which if nothing else
might serve as a useful foil.  If there were heavy pion-like
particles whose decay modes were dominantly 
$\Pi^\pm \rightarrow W + \gamma$ and 
$\Pi^0 \rightarrow ZZ ~{\rm or}~ Z\gamma ~{\rm or}~ \gamma
\gamma$, their production and decay could induce many of the same
final states as we have analyzed above, including the CDF event.  The
missing energy in this event would be carried off by neutrinos; but in
other cases one might have $\gamma \gamma $ without significant
missing energy, so that this alternative will be readily
distinguishable \cite{kolda96}.

These considerations are meant mainly to illustrate how very important
it is, in our present state of ignorance, to obtain some guidance from
experiment.  A few informative events could re-shape
our ideas on particle physics and cosmology
in fundamental ways.  For example, if something like
the minimal messenger model of supersymmetry breaking were established,
we would learn that the lightest R-odd particle is the gravitino, and
that it is
extremely light.  In that case planned experiments to detect a
cosmological background of weakly interacting
massive particles (WIMP) would be doomed to frustration.

\section{Toward Cosmology\protect\cite{brief}}

Now I would like to draw out from this discussion of particle physics
some morals for cosmology.

\bigskip

\noindent$\bullet$ It makes good sense to extrapolate toward the
{\it very\/} early Universe.

As Figure 3 shows us in hard data, the strong coupling
runs toward a small effective value at high energy.  Figure 6, and
especially Figure 7, emphasize why the seductive
assumption that
in crucial respects particle
physics remains simple and weakly coupled up to extraordinarily high
energies is very  hard to resist, because it leads to a strikingly
successful account of the unification of couplings.  Thus fundamental
particle physics becomes profoundly simpler (though
superficially more complex) toward the earliest moments of the Big Bang.

The scale for the running of inverse
couplings, once they are large, is
logarithmic.  This feature naturally connects mass scales identifiable
in particle
physics experiments
with exponentially (in the inverse couplings) larger scales.
Numerically, as we have seen, one is lead in this way close to the
Planck scale.  Thus there is direct,
though of course extremely limited,
evidence that quantum field theory at weak coupling governs the
interactions of the known particles
up to energies (temperatures, densities)
nearly as high as one could reasonably hope.

\bigskip

$\bullet$ Cosmic phase transitions happened.

Since QCD and asymptotic freedom are firmly established,
we can say with complete confidence that the
effective low-energy degrees of freedom in the strong interaction
-- hadrons with confined color and
spontaneously broken chiral symmetry --  are quite different from the
fundamental colored quark and gluon degrees of freedom which manifest
themselves at high temperature.  The transition between them must
be accompanied by a rapid crossover and perhaps by a phase transition.
Indeed, the existence of a phase transition can be rigorously established
in models closely related to real-world QCD, such as the variant of QCD
where the $m_u = m_d = 0$ or the pure-glue theory.  The nature of
the transition, perhaps
surprisingly,
seems to depend sensitively upon the spectrum of light quarks.

The behavior of QCD at high temperatures is in principle, and to some
extent in practice, calculable.  The QCD
crossover or transition that occurred during the Big Bang is even
in some rough sense {\it reproducible}, and will be approximated in
future relativistic heavy ion collisions.

Similarly,
since the electroweak $SU(2)\times U(1)\rightarrow U(1)$ breaking pattern
is firmly established, there is an excellent chance that another phase
transition, associated with $SU(2)\times U(1)$ restoration at
high temperature, occurs.  It is not guaranteed that there is a strict
phase transition, since there is no gauge-invariant order parameter for
the Higgs phase, though
at weak coupling one certainly expects a sharp
transition.  One
of the most interesting projects for future accelerators, which has
have important implications for cosmology,  will
be to map out the relevant parameters so that we can characterize this
crossover or transition.

These `established' examples encourage one to speculate about the
possibility of phase transitions associated with unification, or
perhaps occurring in some hidden sector.

Cosmological
phase transitions have many possible consequences for the history
of the Universe and for observational cosmology, including:

{\it Defects\/} of all sorts, including
textures, strings, and monopoles that could
persist even to the present day.
Truly stable domain walls must be
avoided or inflated away,
but appropriately long-lived ones could be important in
the prior evolution of the Universe.

{\it Inflation}, if one gets trapped in a metastable condition by
a barrier or by weakness of the relevant coupling that drives the
transition.

{\it Gravity waves}, if the scale of the transition is high
or if the transition is sufficiently violent.

{\it Baryogenesis}, under the same qualitative conditions.  Of course,
baryogenesis does not require a phase transition.

It will be fascinating to discover which, if any, of these possibilities is
realized by the electroweak transition.  There is also a great
opportunity, if one can establish  compelling, sufficiently
detailed  models for
unification, or a hidden sector, or ... ,
to examine their cosmological implications.  In particular, as was repeatedly
emphasized in these Dialogues, we need better models of, or alternatives
to, inflation.

\bigskip

$\bullet$ We have specific, credible dark matter candidates, notably
the lightest R-odd particle (LSP) and the axion.

\bigskip
$\bullet$ We have rich material for fantasies.

I will leave to each of you the joy of choosing
his or her own
favorite fantasies from among the many we have heard,
or of inventing new ones.  I would however like briefly to
conclude by alluding a circle of ideas which though it
can be made to sound quite
fantastic
I think is actually deeply implicit in much current thinking
about particle physics and cosmology.

The axion is perhaps the most well-motivated and studied exemplar
of a family of related fields including familons, dilatons, and
moduli fields that in one way or another embody the idea that
what we ordinarily consider `constants' might not be
fundamental parameters fixed in the very formulation of the
laws of Nature, but rather can be considered usefully as dynamical
entities.
In a theory with only one
fundamental, dimensional, parameter, such as superstring theory appears
to be, there is clearly a sense in which all dimensionless
couplings are
dynamical variables.
They might nevertheless behave effectively
as constants either if it costs a very large energy-density to excite
them at all ({\it e.g}. massive fields)
or if ordinary matter couples very weakly
to long-wavelength fluctuations of the field ({\it e.g}.
stiff light fields with
derivative couplings).

In the latter case one might anticipate
long-range forces mediated by the exchange of the field \cite{moody84}.
I believe
that experiments to look for such forces are among the most fundamental
which can presently
be attempted.  They address in a concrete way the question: are the
constants of Nature uniquely determined to be what we observe by dynamical
laws, or are they `frozen accidents' imprinted at the Big Bang?  Or are
they presently relaxing towards some more favorable
value\footnote{The standard
picture of axion production in the Big Bang is essentially
an example of this: the axion
field starts drops out of equilibrium as the Universe cools from
$10^{12}$ to about 1 Gev, then relaxes towards the dynamically favored,
approximately P and T conserving, value.}?  Note that although a
rapidly (on cosmological time scales) oscillating field represents
non-relativistic matter, a sufficiently slowly varying field
might appear as a contribution to the effective cosmological term.

If these ideas are along the right lines, it could be misguided to seek
a unique Lorentz-invariant `vacuum' state as a model for our present world.
One might instead be required, at a fundamental level, to seek the boundary
conditions for particle physics from cosmology (and {\it vice versa\/}).
One could then hope that
some single simple, attractive {\it ansatz\/} would form the basis of both
cosmology and particle physics, and truly unify these fundamental sciences.
Is it plausible?

I think that recent developments in string theory,
especially in its application to the theory of black holes, are quite
encouraging in this regard.  In that context one primarily deals with
spatially inhomogeneous but essentially stationary (in time) configurations.
Great insight has been achieved by modeling extremal black holes and other
solitons as extended singular surfaces (D-branes) \cite{polchinski96} 
where the conditions for
string propagation are modified.
A central point worth emphasizing that there is no
obvious singularity in directly physical quantities, because there are no
pointlike physical objects within the theory capable of
probing and
resolving the singular mathematical structure directly.  At  large distances
one has physics which appears to a local observer to be building toward
a field-theoretic soliton and spatial geometry with a singular core, but
the crisis dissolves into something that cannot be described in terms
of ordinary fields and spatial geometry.
It ought to be possible to examine the analogous situation for
pseudo-singularities occuring at an initial `time' -- which of course would
wind up {\it not\/} being physical time.  In this way, it might be possible
to retain the advantages of simplicity and uniqueness inherent in
a mathematically singular initial condition, while avoiding any associated
physical pathology.

It is conceivable to me that even a very simple initial condition might
suffice to generate a region of the
Universe of sufficient  complexity to be
identified with our own.  Indeed, by analogy with the
black hole case one might expect that the (mathematically) singular
time-origin would look, at sufficient distance so that familiar
space-time-matter notions
apply, like a sort of Universal heat pulse -- which perhaps could be
identified as the trigger
for a hot Big Bang.  As Guth and Linde have emphasized to us,
such a quasi-thermal starting point can lead, through inflation, to an
extremely rich and structured dynamics, including generically the possibility
that different spatial regions are governed by different `laws of physics'
-- specifically differing with respect to
the values of the sorts of moduli fields
mentioned above,
but also conceivably differing even with respect to effective
spatial dimension, gauge groups and family
structures, supersymmetry breaking (or not) mechanism, and so forth.
Physics would then give an essentially historical -- but not vague --
account of the world we observe: an account highly constrained
but perhaps far from determined
by fundamental symmetry and consistency principles.  This would be
similar to how
biology is constrained, but not
determined, by physics.

Of course, this meta-proposal for a wave function -- or rather
initial condition  -- of the Universe is based on quite different physical
considerations from that of Hartle-Hawking \cite{hartle83}, 
and probably contradicts it:
in particular
there is no appeal to Euclidean procedures,
and the arrow of time is manifest.

\bigskip

\bigskip

My overarching perception, coming out of this set of Dialogues, is
that the doors to several new worlds are ajar.  On the cosmological side,
we can look forward especially
to new observational results on the microwave anisotropy.
These will heavily impact the classic questions of homogeneous cosmology
(including
the values of Hubble parameter $h$, the spatial curvature of the Universe,
and the cosmological term $\lambda$) as well as theoretical ideas
about primordial fluctuations.
The
semi-phenomenological approach to particle physics, emphasized in
the body of this talk, leads us to anticipate
that there is a wealth of accessible new phenomena
whose broad outlines (I believe) we
can confidently foresee,
and whose details will be richly informative on fundamental
questions.  And there is today rapid ferment in the `high theory'
of particle physics and quantum gravity, with results of striking
novelty and mathematical depth emerging.  The challenge remains,
to connect these results with concrete phenomena.  Perhaps cosmology
offers our best hope for this.

We see narrow shafts
of light emerging from behind each door.  In each case, we
do not know what is on
the other side -- whether it is a beautiful habitable world,
a labyrinth, or a jungle.
The doors may be heavy, and require elaborate machinery
to move them, but we must, and will, push through.

\section*{Acknowledgments}
I  would like to thank Keith Dienes for supplying
Figures 6 and 7, and especially Chris Kolda for
valuable assistance in the
preparation of this manuscript.


\noindent

\begin{thebibliography}{99}

\bibitem{sm}
For recent reviews of the Standard Model, see References 4 and 6.

\bibitem{weinsalam}
After several partial and tentative proposals, the $SU(2)\times U(1)$
electroweak theory took on in its essentially modern form in:
S. Weinberg, Phys. Rev. Lett. {\bf 19}, 1264 (1967); 
A. Salam, in {\it Elementary Particle
Physics}, ed. N. Svartholm (Almqvist and Wiksells, Stockholm, 1968), p. 367;
S. Glashow, J. Iliopoulos, and L. Maiani,  Phys. Rev. D{\bf 2}, 1285
(1970).

\bibitem{sutheory}
After several partial and tentative proposals, 
the $SU(3)$ strong
interaction theory took on its essentially modern form in:  
D. Gross and F. Wilczek, Phys Rev. D {\bf 8}, 3633 (1973);
S. Weinberg, Phys. Rev. Lett. {\bf 31}, 494 (1973);
H. Fritzsch, M. Gell-Mann, and H. Leutwyler,  Phys. Lett. {\bf 47B}, 
365 (1973).  Not coincidentally,
the key discovery that allowed one to connect the abstract
gauge theory to
experiments is asymptotic freedom was first demonstrated
just prior
to these papers, in:
D.~Gross and F. Wilczek, Phys. Rev. Lett. {\bf 30}, 1343 (1973);
H. D. Politzer, Phys. Rev. Lett. {\bf 30}, 1346 (1973).

\bibitem{ewfig}
LEP Electroweak Working Group, report LEPEWWG/95-02 (Aug 1995).

\bibitem{bkm}
K.S. Babu, C. Kolda and J. March-Russell, IASSNS-HEP-96/20, hep-ph/9603212
(Feb 1996), Phys. Rev D. (in press).

\bibitem{qcdfig}
M. Schmelling, talk given at the XV International Conference on Physics in 
Collision, Cracow, Poland, June 1995, CERN-PPE/95-29 (Aug 1995).

\bibitem{hooft}
G. 't Hooft, Phys. Rev. Lett. {\bf 37}, 8 (1976); C. Callan, R. Dashen,
and D. Gross, Phys. Lett. {\bf 63B}, 334 (1976); R. Jackiw, C. Rebbi, Phys.
Rev. Lett. {\bf 37}, 172 (1976).

\bibitem{pecci77}
R. Peccei and H. Quinn, Phys. Rev. Lett. {\bf 38}, 1440
(1977),  Phys. Rev. D {\bf 16}, 1791 (1977).

\bibitem{weinberg78}
S. Weinberg, Phys. Rev. Lett. {\bf 40}, 223 (1978); F. Wilczek, 
Phys. Rev. Lett. {\bf 40}, 279 (1978).

\bibitem{preskill83}
J. Preskill, M. Wise, and F. Wilczek, Phys. Lett.
{\bf 120B}, 127 (1983);
L. Abbott and P. Sikivie, Phys. Lett. {\bf 120B}, 133 (1983); M. Dine and
W. Fischler, Phys. Lett. {\bf 120B}, 137 (1983).

\bibitem{topquark}
The existence of the infrared fixed point was first discussed in:
C.~Hill, Phys. Rev. D {\bf 24},  691 (1981). More recent examinations 
including supersymmetry appear in:    V. Barger, M. Berger and P.
Ohmann, Phys. Rev. D {\bf 47}, 1093 (1993);  P. Langacker and N. 
Polonsky, Phys. Rev. D {\bf 47}, 4028 (1993); M. Carena, S.~Pokorski 
and C.~Wagner, Nuc. Phys. B {\bf 406}, 59 (1993).

\bibitem{jarlskog85}
C. Jarlskog,  Phys. Rev. Lett. {\bf 55}, 1039 (1985). 

\bibitem{bahcall95}J. Bahcall, {\it et al}, Nature {\bf 375}, 29 
(1995).

\bibitem{lisi94}
G. Fogli, E. Lisi, and D. Montanino, Phys. Rev. D {\bf 49}, 3626 (1994). 

\bibitem{athan95}
C. Athanassopoulos, {\it et al}, Phys. Rev. Lett.
{\bf 75}, 2650 (1995).

\bibitem{gell-mann79}
M. Gell-Mann, P. Ramond, and R. Slansky, in 
{\it Supergravity}, ed. P. van Neiuwenhuizen and D. Freedman 
(North Holland, Amsterdam, 1979),
p. 315; T.~Yanagida, Proc. of the Workshop on Unified Theory and
Baryon Number in the Universe, eds. O. Sawada and A. Sugamoto (KEK, 1979).

\bibitem{hawking76}
S. Hawking, Phys. Rev. D {\bf 14}, 2460 (1976).

\bibitem{georgi74}
H. Georgi and S. Glashow, Phys. Rev. Lett. {\bf 32}, 438 (1974).

\bibitem{georgi75}
H. Georgi, in {\it Particles and Fields -- 1974}, ed. C. Carlson
(AIP press, New York, 1975).

\bibitem{unification}
S.~Dimopoulos, S.~Raby and F.~Wilczek, Phys. Today {\bf44}, 25 (1991) 
and references contained therein.

\bibitem{nielsen81}
N. Nielsen, Am. J. Phys. {\bf 49}, 1171 (1981); R. Hughes, Nucl. Phys. B
{\bf 186}, 376 (1981).

\bibitem{quinn74}
H. Georgi, H. Quinn, and S. Weinberg, Phys. Rev. Lett. {\bf 33}, 451 (1974).

\bibitem{blewitt85}
See for example G. Blewitt, {\it et al}, Phys. Rev. Lett. {\bf 55}, 
2114 (1985), and 
the latest Particle Data Group compilations.

\bibitem{ferrara86}
A very useful introduction and collection of basic papers
on supersymmetry
is S. Ferrara, {\it Supersymmetry\/} (2 vols.) (World Scientific,
Singapore 1986).   Another excellent standard reference
is N.-P. Nilles, Phys. Reports {\bf 110}, 1 (1984).

\bibitem{dimopoulos81}S. Dimopoulos, S. Raby, and F.~Wilczek, Phys.
Rev. D {\bf 24}, 1681 (1981).

\bibitem{ellis91}
J. Ellis, S. Kelley, and D. Nanopoulos,
Phys. Lett. {\bf B260}, 131 (1991);
U. Amaldi, W. de Boer, and H. Furstenau, 
Phys. Lett. {\bf B260}, 447 (1991); for more recent analysis see
P. Langacker and N. Polonsky, Phys. Rev. D {\bf 49}, 1454 (1994).

\bibitem{dienes96}
K. Dienes, IASSNS-HEP-95/97, hep-th/9602045 (Feb 1996), Phys. Reports  
(in press).

\bibitem{kolda96}
This section is essentially a condensed version of
K. S. Babu, C.~Kolda, and F.~Wilczek, IASSNS-HEP96/55, hep-ph/9605408
(May 1996).

\bibitem{models}
M.~Dine and A.~Nelson, Phys. Rev. D {\bf 48}, 1277 (1993); 
M.~Dine, A.~Nelson and Y.~Shirman, Phys. Rev. D {\bf 51}, 1362 (1995); 
M.~Dine, A.~Nelson, Y.~Nir and Y.~Shirman, Phys. Rev. D {\bf 53}, 2658 (1996).

\bibitem{fayet}
P.~Fayet, Physics Reports {\bf 105}, 21 (1984).

\bibitem{signals}
S.~Dimopoulos, M.~Dine, S.~Raby and S.~Thomas, Phys. Rev. Lett. {\bf76}, 
3494 (1996); 
S.~Ambrosanio, G.~Kane, G.~Kribs, S.~Martin and S.~Mrenna,
Phys.  Rev. Lett. {\bf 76}, 3498 (1996); 
S.~Dimopoulos, S.~Thomas, and
J.~Wells, report SLAC-PUB-7148, hep-ph/9604452 (April 1996); 
G.~Dvali, G.~Giudice and A.~Pomarol, report CERN-TH/96-61, 
hep-ph/9603238 (March 1996).

\bibitem{cdf}S.~Park, representing the CDF Collaboration, ``Search for New
Phenomena in CDF,'' in {\sl 10th Topical Workshop on Proton-Antiproton
Collider Physics}, ed. by R.~Raha and J.~Yoh (AIP Press, New York, 
1995), report FERMILAB-CONF-95/155-E.

\bibitem{brief}
Since in this Section many large
ideas will be mentioned very briefly, I will
not attempt systematic citations.

\bibitem{moody84}
J. Moody and F. Wilczek, Phys. Rev. D {\bf 30}, 130 (1984).

\bibitem{polchinski96}
J. Polchinski, S. Chaudhuri and C. Johnson, NSF-ITP-96-003, 
hep-th/9602052 (January 1996).

\bibitem{hartle83}
J. Hartle and S. Hawking, Phys. Rev. D {\bf 28}, 2960 (1983).

\end{thebibliography}
\end{document}